\documentclass[aps,prl,reprint,showpacs,floatfix]{revtex4-1}
\usepackage{amsmath,graphics,epsfig,color,verbatim}

\let\oldhat\hat
\renewcommand{\vec}[1]{\mathbf{#1}}
\renewcommand{\hat}[1]{\oldhat{\mathbf{#1}}}

\begin{document}

\title{Ideal strength and phonon instability of strained monolayer materials}
\author{Eric B. Isaacs}
\email{ebi2104@columbia.edu}
\author{Chris A. Marianetti}
\email{chris.marianetti@columbia.edu}
\affiliation{Department of Applied Physics and Applied Mathematics, Columbia University, New York, NY 10027}

\begin{abstract}
The ideal strength of monolayer materials possessing semimetallic,
semiconducting, and insulating ground states is computed using density
functional theory. Here we show that, as in graphene, a soft mode
occurs at the K-point in BN, graphane, and MoS$_2$, while not in
silicene. The transition is first-order in all cases except graphene.
In BN and graphane the soft mode corresponds to a Kekul{\'e}-like
distortion similar to that of graphene, while MoS$_2$ has a distinct
distortion. The phase transitions for BN, graphane, and MoS$_2$ are
not associated with the opening of a band gap, which indicates that
Fermi surface nesting is not the driving force. We perform an energy
decomposition that demonstrates why the soft modes at the K-point are
unique and how strain drives the phonon instability.
\end{abstract}

\date{\today}
\maketitle

Ideal strength, the maximum stress an infinite, defect-free crystal
can withstand at zero temperature, is an upper limit that provides a
measure for the intrinsic strength of the chemical bonding and overall
stability of a material \cite{kelly1986strong}. Ideal strength is
ultimately dictated by what is known as the \textit{elastic
  instability}, whereby a crystal becomes unstable with respect to a
homogeneous deformation along the strain path. This scenario
corresponds to an imaginary-frequency or ``soft'' phonon mode of
vanishing wavevector ($q\rightarrow0$) and a maximum in the
stress-strain curve. However, a finite-wavevector phonon instability,
known as a \textit{soft mode}, occurring at a lower stress than that
of the elastic instability can also limit a material's ideal strength
via the transformation to a new structure with a lower elastic
instability. Acoustic phonon instabilities have been predicted to
limit the ideal strength of bulk aluminum
\cite{clatterbuck_phonon_2003} and bulk silicon
\cite{dubois_ideal_2006} for certain strain modes.

Monolayer materials are an optimal testbed for studying the
possibility of strength-limiting soft modes since they can be
fabricated with unprecedented levels of crystalline perfection. Under
conditions at or close to equibiaxial strain, the mechanical failure
of graphene was found to stem from an optical phonon instability at
the K-point of the Brillouin zone (BZ) in which the pristine honeycomb
structure distorts towards a Kekul{\'e}-like structure of isolated
C$_6$ regular hexagonal rings \cite{marianetti_failure_2010}. Since
this structural transformation gaps the Fermi surface by breaking the
symmetry of the honeycomb structure
\cite{semenoff_condensed-matter_1984,hou_electron_2007}, it has been
proposed that the soft mode in graphene is a two-dimensional (2D)
manifestation of a Peierls instability \cite{lee_band_2011}. This has
stimulated work documenting the effect of doping on the instability
\cite{si_electronic_2012,woo_ideal_2013}. Given that the essence of
the Peierls instability arises from the properties of one-dimensional
systems, clearly this analogy is strained and this instability cannot
solely be attributed to the Fermi surface. Nonetheless, the degree to
which the gapping of the Dirac point drives the instability is an open
question. Li recently found that single-layer molybdenum disulfide
(MoS$_2$) also exhibits a soft mode under equibiaxial strain
\cite{li_ideal_2012}, which further raises the question of the origin
and generality of phonon instabilities in monolayer materials.

Here we employ density functional theory (DFT) calculations to
investigate a structurally and electronically diverse set of existing
2D crystals---graphene, single-layer boron nitride (BN), graphane,
MoS$_2$, and silicene---under equibiaxial strain in order to gain
insight into the nature of phonon instabilities in monolayer
materials. In addition to graphene and MoS$_2$, we find a soft mode at
the K-point for BN and graphane leading to mechanical failure for
BN. We show that the nature of the distortion in BN is completely
analogous to graphene, despite the fact that BN has a large band
gap. This illustrates that the Fermi surface is not the general
driving force of this instability. In order to elucidate the physics
of this instability, we perform a decomposition of the total energy
into two terms which reasonably embody the electronic and elastic
aspects of the energetics. This demonstrates the potency of the
electronic term for the K-point soft mode in addition to the rapid
decay of the elastic term as a function of strain.

Non-spin-polarized DFT
\cite{hohenberg_inhomogeneous_1964,kohn_self-consistent_1965}
calculations within the generalized gradient approximation of Perdew,
Burke, and Ernzerhof \cite{perdew_generalized_1996} are performed
using the Vienna \textit{ab initio} simulation package (\textsc{vasp})
\cite{kresse_ab_1994,kresse_ab_1993,kresse_efficient_1996,kresse_efficiency_1996}.
The Kohn-Sham equations are solved using a plane-wave basis set
(kinetic energy cutoff of 420 eV for MoS$_2$ and silicene, 450 eV for
graphene and graphane, and 500 eV for BN) and the projector augmented
wave method \cite{blochl_projector_1994,kresse_ultrasoft_1999} with
soft projectors for B, C, and N. The primitive unit cell in-plane
lattice vectors are chosen to be
$\vec{a_1}=\sqrt{3}l/2\hspace{1mm}\hat{x}-3l/2\hspace{1mm}\hat{y}$ and
$\vec{a_2}=\sqrt{3}l/2\hspace{1mm}\hat{x}+3l/2\hspace{1mm}\hat{y}$,
where $l$ is the in-plane length of the nearest-neighbor C-C, B-N,
Mo-S, and Si-Si bond for graphene and graphane, BN, MoS$_2$, and
silicene, respectively. The in-plane lattice vectors of the K-cell
supercell \cite{marianetti_failure_2010} commensurate with a K-point
lattice distortion are $\vec{A_1}=2\vec{a_1}+\vec{a_2}$ and
$\vec{A_2}=\vec{a_1}+2\vec{a_2}$. The out-of-plane lattice vector
length is chosen to be 14\ \AA\ for graphene, BN, graphane, and
silicene and 16\ \AA\ for MoS$_2$. To sample reciprocal space we
employ $k$-point grids of $20\times20\times1$ for MoS$_2$ and silicene
and $24\times24\times1$ for graphene, graphane, and BN for the
primitive cell and $8\times8\times1$ for graphene, BN, and MoS$_2$ and
$9\times9\times1$ for graphane for the K-cell. The total energy, ionic
positions, and stress tensor components are converged to 10$^{-6}$ eV,
0.01 eV/\AA, and 10$^{-3}$ GPa, respectively. Phonons at the K-point
are obtained using the frozen phonon method. To compute stress-strain
curves the unit cell is equibiaxially strained, ionic positions are
perturbed to allow symmetry breaking, and then the ions are fully
relaxed. We renormalize the equibiaxial true stress
$\sigma=(\sigma_{xx}+\sigma_{yy})/\sqrt{2}$ of each 2D material to the
interlayer spacing of the most closely related bulk
material \footnote{We choose 3.35 \AA\ \cite{baskin_lattice_1955},
  3.33 \AA\ \cite{liu_structural_2003}, and 6.15
  \AA\ \cite{boker_band_2001} for graphene and graphane, BN, and
  MoS$_2$, respectively.} to give a physical reference for stress
values. Density functional perturbation theory
\cite{baroni_phonons_2001} calculations in the \textsc{quantum
  espresso} package \cite{giannozzi_quantum_2009} are performed with a
$10\times10\times1$ $q$-point grid for the initial search for soft
modes as a function of equibiaxial strain.

\begin{figure}[t]
\begin{center}
\includegraphics[width=\linewidth]{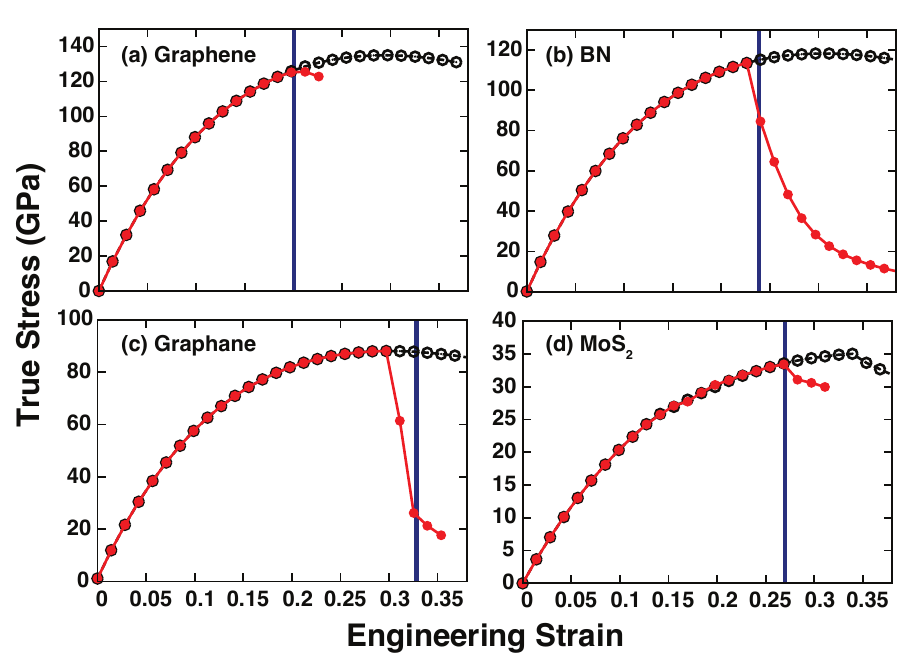}
\end{center}
\caption{ True stress $\sigma$ vs. engineering strain $\varepsilon$
  for \textbf{(a)\hspace{1mm}} graphene, \textbf{(b)\hspace{1mm}} BN,
  \textbf{(c)\hspace{1mm}} graphane, and \textbf{(d)\hspace{1mm}}
  MoS$_2$ under equibiaxial strain. Black lines and open circles for
  the primitive unit cell; red lines and filled circles for the
  K-cell. The strain at which a phonon mode goes soft at the K-point
  is indicated by a blue line.
\label{stress}
}
\end{figure}

For BN and graphane, in addition to graphene
\cite{marianetti_failure_2010} and MoS$_2$ \cite{li_ideal_2012}, under
equibiaxial strain the first instance of the eigenvalues of a phonon
branch becoming imaginary at a finite wavevector occurs at the
K-point. The critical values of equibiaxial engineering strain
$\varepsilon=(\varepsilon_{xx}+\varepsilon_{yy})/\sqrt{2}$ at which
the phonon mode goes soft at the K-point computed via the frozen
phonon method are 0.201, 0.239, 0.328, and 0.270 for graphene, BN,
graphane, and MoS$_2$, respectively. No finite-wavevector soft modes
preceding the elastic instability are found for silicene.

\begin{figure}[b]
\begin{center}
\includegraphics[width=\linewidth]{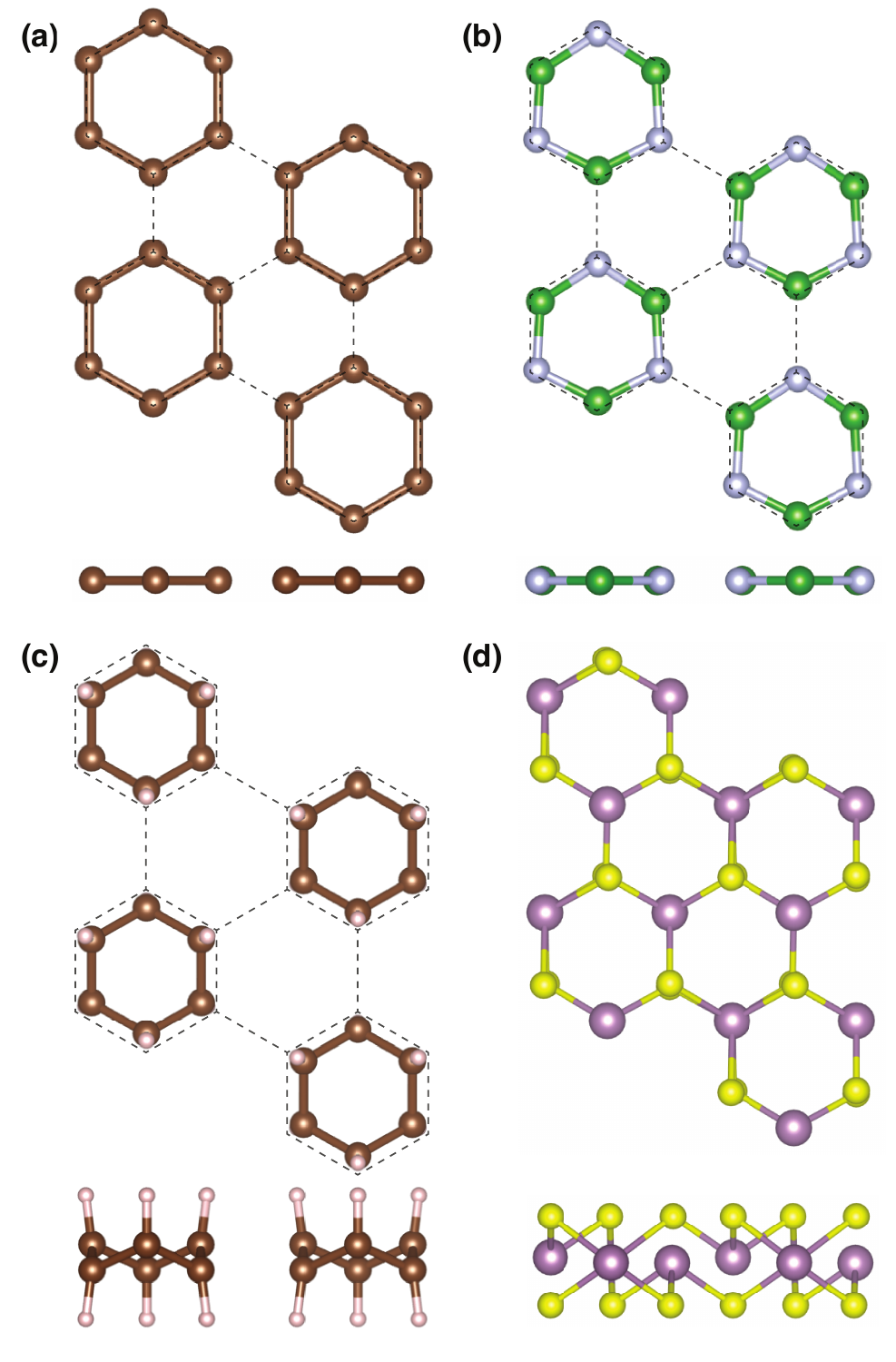}
\end{center}
\caption{Top and side orthographic projections of the distorted
  structures for \textbf{(a)\hspace{1mm}} graphene,
  \textbf{(b)\hspace{1mm}} BN, \textbf{(c)\hspace{1mm}} graphane, and
  \textbf{(d)\hspace{1mm}} MoS$_2$ at equibiaxial strains of 0.212,
  0.240, 0.328, and 0.270, respectively. The C, B, N, H, Mo, and S
  atoms are represented as brown, green, silver, white, purple, and
  yellow spheres, respectively. Dashed lines indicate the undistorted
  strained lattice.
\label{structures}
}
\end{figure}

\begin{figure*}[t]
\includegraphics[width=\textwidth]{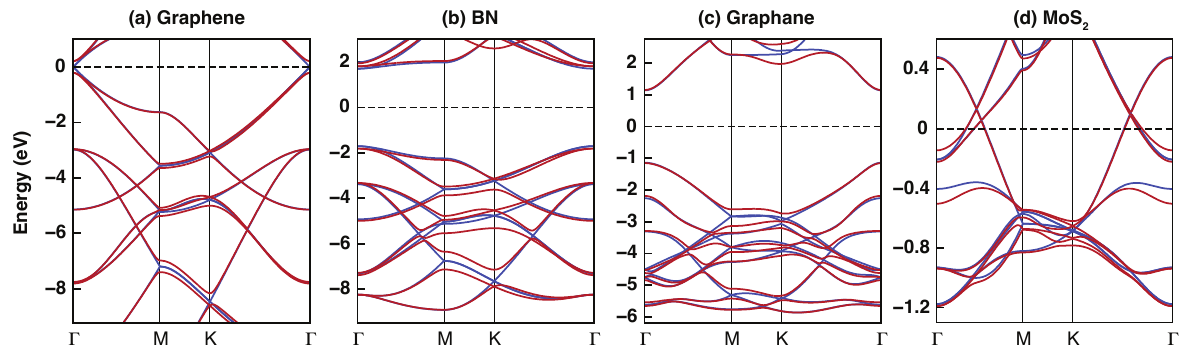}
\caption{K-cell electronic band structures for
  \textbf{(a)\hspace{1mm}} graphene, \textbf{(b)\hspace{1mm}} BN,
  \textbf{(c)\hspace{1mm}} graphane, and \textbf{(d)\hspace{1mm}}
  MoS$_2$ at equibiaxial strains of 0.212, 0.240, 0.328, 0.270,
  respectively. Blue lines for the undistorted structure; red lines
  for the partially-distorted structure corresponding to a 50\%/50\%,
  60\%/40\%, 90\%/10\%, and 25\%/75\% linear combination of the
  undistorted/distorted structures for graphene, BN, graphane, and
  MoS$_2$, respectively. The $k$-point labels $\Gamma$, M, and K
  correspond to the center, edge midpoint, and corner of the BZ and
  the dashed black line indicates the band gap midpoint for insulators
  and the Fermi energy for MoS$_2$.
\label{bands_vs_amp}
}
\end{figure*}

To explore the impact of the K-point soft mode on the ideal strength,
in Fig. \ref{stress} we compare the stress-strain curve of the K-cell
commensurate with a K-point lattice distortion to that of the
primitive cell. At critical values of strain identical or close to
those found via the frozen phonon method, the K-cell curves shown in
red significantly deviate from the primitive cell curves shown in
black in the form of a drop in stress associated with a transformation
to a new structure with a lower elastic instability. Computing the
stress-strain curve on a finer grid of strain values near this
transformation and tracking the changes in the relaxed ionic positions
reveals that while the phase transition of graphene is continuous, BN,
graphane, and MoS$_2$ each undergo a first-order phase transition with
a sharp discontinuity in the stress and bond lengths. The first-order
nature is most apparent in graphane, for which the distorted structure
becomes the ground state noticeably before the phonon goes soft. The
elastic instability, corresponding to the peak of the primitive cell
curves, occurs at a strain (stress) of 0.297 (135.0 GPa), 0.311 (118.2
GPa), 0.297 (88.0 GPa), and 0.339 (35.0 GPa) for graphene, BN,
graphane, and MoS$_2$, respectively. The ideal strength of graphene,
BN, and MoS$_2$ are limited by the phonon instabilities since they
correspond to substantially reduced strain (stress) values of 0.206
(125.9 GPa), 0.231 (114.1 GPa), and 0.269 (33.4 GPa), respectively. In
contrast, for graphane the phonon instability does not precede the
elastic instability so we do not predict the ideal strength is reduced
by the K-point soft mode.

The distorted structures that result from the soft modes are
illustrated in Fig. \ref{structures}. Like in the case of graphene,
the soft mode has a 2D irreducible representation and anharmonicity
determines the minimum-energy direction and hence the ground-state
structure \cite{marianetti_failure_2010}. Graphene, BN, and graphane
distort towards Kekul{\'e}-like structures consisting of isolated
units of C$_6$ regular hexagons for graphene, B$_3$N$_3$ irregular
hexagons for BN, and buckled C$_6$H$_6$ structures similar to that of
the chair conformation of cyclohexane (without the equatorial H atoms)
for graphane. Such distortions have a beautiful classical analogy in
strained porous elastomeric sheets, whose failure modes under
equibiaxial strain correspond to arrays of alternating smaller and
larger pores \cite{michel_microscopic_2007}. MoS$_2$ undergoes a
distinct structural transformation in which Mo and S atoms move
out-of- and in-plane, respectively. In the distorted structure one of
the three Mo sites has six nearly-equal Mo-S nearest-neighbor bond
lengths, and two of the three Mo sites distort (one in the $+\hat{z}$
direction and one in the $-\hat{z}$ direction, where $\hat{z}$ is the
out-of-plane direction) towards trigonal pyramidal coordination with
three nearest-neighbor S atoms.

To investigate the nature and mechanism of the phonon instabilities,
in Fig. \ref{bands_vs_amp} we examine the K-cell electronic band
structures with and without distortion at critical strain, i.e.,
strained at or just beyond the onset of the soft phonon mode. The
particular amounts of distortion, which correspond to a 50\%/50\%,
60\%/40\%, 90\%/10\%, and 25\%/75\% linear combination of the
undistorted/distorted structures for graphene, BN, graphane, and
MoS$_2$, respectively, are chosen to most clearly illustrate how the
soft mode affects the electronic bands. For graphene
(Fig. \ref{bands_vs_amp}a) a gap opens at the $\Gamma$-point,
corresponding to the K-point of the primitive cell due to zone
folding, consistent with the Peierls instability picture. However,
there are also numerous nonlinear splittings of degenerate bands at
lower energy in graphene as well as in BN, graphane, and MoS$_2$. The
structural distortions tend to break degeneracies and disentangle
groups of bands. In some cases, such as in MoS$_2$
(Fig. \ref{bands_vs_amp}d), specific bands substantially shift towards
lower energy in parts of the BZ. BN (Fig. \ref{bands_vs_amp}b) and
graphane (Fig. \ref{bands_vs_amp}c) are insulating in the undistorted
state with substantial gaps of 3.4 eV and 3.3 eV, respectively. While
for BN the distorted structure remains insulating, for graphane after
the onset of the phonon instability the fully distorted structure
(whose bands are not shown) is semimetallic. For MoS$_2$, a
semiconductor in its unstrained state that becomes semimetallic at an
equibiaxial strain of approximately 0.13
\cite{li_ideal_2012,scalise_strain-induced_2012}, the structural
distortion does not open up a gap as indicated by the multiple bands
passing through the Fermi energy for the partially-distorted
structure. Since the soft mode distortion is not accompanied by a band
gap opening for BN, graphane, and MoS$_2$, it is clear that a 2D
analogy to the Peierls distortion cannot be the underlying mechanism
in general. Furthermore, the fact that BN and graphane have
substantial band gaps and exhibit very similar soft modes to that of
graphene strongly suggests that for graphene the gapping of the Dirac
point is more of a consequence than a cause of the phonon instability.

In order to further elucidate the mechanism of the phonon
instabilities and to quantitatively examine different effects, we
introduce a scheme to partition the total energy in order to compare
the two most electronically disparate cases: graphene (a semimetal)
and BN (an insulator with a large gap). The total energy expression in
DFT can be written as a function of the Kohn-Sham eigenvalues:
\begin{align}\label{} \nonumber
E_{tot}&=\sum_{i,\vec{k}}\epsilon_{i\vec{k}}\theta(\epsilon_{F}-\epsilon_{i\vec{k}})-\frac{1}{2}\int \frac{\rho(\vec{r})\rho(\vec{r'})}{|\vec{r}-\vec{r'}|}\,d^3r\,d^3r' \\
& +E_{xc}[\rho] -\int v_{xc}(\vec{r})\rho(\vec{r})\,d^3r+E_{nuc}
\end{align}
where $\epsilon_{i\vec{k}}$ is the Kohn-Sham eigenvalue of the $i$th
band at $k$-point $\vec{k}$, $\epsilon_F$ is the Fermi energy,
$\rho(\vec{r})$ is the charge density at position $\vec{r}$, $E_{xc}$
and $v_{xc}$ are the exchange-correlation energy and potential,
respectively, and $E_{nuc}$ is the electrostatic energy of the
nuclei. In the spirit of previous work on Fermi surface nesting
\cite{mintmire_local-density-functional_1987,ashkenazi_ground_1989},
we partition $E_{tot}$ into an ``electronic'' band energy $E_{elec}$
and an ``elastic'' energy $E_{elas}$ defined as follows:
\begin{align}\label{}
E_{elec}&=\sum_{i,\vec{k}}(\epsilon_{i\vec{k}}-\epsilon_a)\theta(\epsilon_{F}-\epsilon_{i\vec{k}}) \\
E_{elas}&=E_{tot}-E_{elec},
\end{align}
where a reference ``anchor'' $\epsilon_a$ from which to measure the
Kohn-Sham eigenvalues is necessary since there is an arbitrary
constant in the treatment of an infinite interacting system related to
individually divergent summations \cite{Martin2008}. For the anchor we
choose the average of the highest occupied and lowest unoccupied
eigenvalues, e.g. the Dirac point states for graphene, which remain
stationary to first order for a non-interacting system. The band
structures in Fig. \ref{bands_vs_amp}a-c are plotted with respect to
this anchor choice.
\begin{figure}[h]
\includegraphics[width=\linewidth]{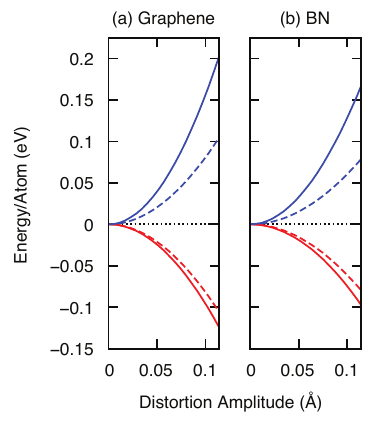}
\caption{Change in electronic (red) and elastic (blue) energies as a
  function of soft mode distortion amplitude for the K-cell structures
  of \textbf{(a)\hspace{1mm}} graphene and \textbf{(b)\hspace{1mm}} BN
  at equibiaxial strains (dashed lines) of 0.212 and 0.240,
  respectively, compared to that at zero strain (solid lines).
\label{energy_partition}
}
\end{figure}

In Fig. \ref{energy_partition} we plot $E_{elec}$ and $E_{elas}$ as a
function of the soft mode distortion amplitude for unstrained and
critically-strained graphene and BN. It should be noted that only the
quadratic regime is relevant in terms of deducing the instability. For
graphene (Fig. \ref{energy_partition}a) and BN
(Fig. \ref{energy_partition}b) $E_{elec}$ is negative, indicating that
changes in band energy drive the soft mode transitions. Since BN is
insulating, this demonstrates that band energy lowering can be
appreciable in such phase transitions even in the absence of a Fermi
surface \cite{johannes_fermi_2008}. For graphene we performed this
partition for all the other modes at the K-point in addition to
selected modes at the M-point (not shown), and no other mode had such
a large, negative quadratic coefficient. Every negative electronic
term was at least 2--3 times smaller in magnitude. For both graphene
and BN the magnitude of the band energy lowering decreases with
strain, and therefore strain actually \textit{weakens} the electronic
driving force despite the fact that it is essential for triggering the
transition. The phonon instability emerges since $E_{elas}$ decays
much more rapidly as a function of strain. Therefore, the key role of
strain is to soften the elastic term such that the electronic term can
dominate and drive the total energy negative. For both graphene and BN
it is this strain-induced softening of the elastic term that enables
the soft mode. Important future work will be building a physical
understanding of how and why particular modes at the K-point have such
a strong electronic term.


In conclusion, using DFT calculations we find soft modes similar to
that of graphene for BN, graphane, and MoS$_2$ that limit the ideal
strength of BN and MoS$_2$ under equibiaxial strain. While for BN and
graphane the soft mode corresponds to a Kekul{\'e}-like distortion
similar to that of graphene, MoS$_2$ has a distinct soft mode in which
2/3 of the Mo sites distort towards trigonal pyramidal
coordination. The structural transitions for BN, graphane, and MoS$_2$
are not associated with the opening of a band gap, which reveals that
Fermi surface nesting does not generally play a role in these
transitions. Decomposing the total energy elucidates the complementary
roles of a large band energy lowering that decays slowly with strain
and a rapidly-decaying elastic energy penalty in driving phonon
instabilities in monolayer materials.

\begin{acknowledgments}
The authors acknowledge support from the National Science Foundation
(Grant No. CMMI-0927891) and the New York Center for Computational
Sciences at Stony Brook University/Brookhaven National Laboratory
which is supported by the U.S. Department of Energy (Grant
No. DE-AC02-98CH10886) and the State of New York. E.B.I. gratefully
acknowledges financial support from the U.S. Department of Energy
Computational Science Graduate Fellowship (Grant
No. DE-FG02-97ER25308).
\end{acknowledgments}

\bibliography{../references/main}

\end{document}